# Description of the Early Growth Dynamics of 2014 West Africa Ebola Epidemic

Maria A. Kiskowski
Department of Mathematics and Statistics,
University of South Alabama

## Abstract

Background: The early growth dynamics of the West African Ebola virus epidemic has been qualitatively different for Guinea, Sierra Leone and Liberia. However, it is important to understand these disparate dynamics as trends of a single epidemic spread over regions with similar geographic and cultural aspects, with likely common parameters for transmission rates and the reproduction number $R_0$.

Methods: We combine a discrete, stochastic SEIR model with a two-scale community network model to demonstrate that the different regional trends may be explained by different community mixing rates. Heuristically, the effect of different community mixing rates may be understood as the observation that two individuals infected by the same chain of transmission are more likely to know one another in a less-mixed community. Saturation effects occur as the contacts of an infected individual are more likely to already be exposed by the same chain of transmission.

Results: The effects of community mixing, together with the effects of stochasticity, can explain the qualitative difference in the growth of Ebola virus cases in each country, and why the probability of large outbreaks may have recently increased. An increase in the rate of Ebola cases in Guinea in late August, and a local fitting of the transient dynamics of the Ebola cases in Liberia, suggests that the epidemic in Liberia has been more severe, and the epidemic in Guinea is worsening, due to discrete seeding events as the epidemic spreads into new communities.

Conclusions: A relatively simple network model provides insight on the role of local effects such as saturation that would be difficult to otherwise quantify. Our results predict that exponential growth of an epidemic is driven by the exposure of new communities, underscoring the importance of control measures that limit this spread.

# Introduction

As of mid-October 2014, the number of reported suspected cases of the Ebola epidemic in West Africa had exceeded 9,000 cases, which is likely a significant underestimate ( [1, 2, 3]). Markedly different dynamical behaviors can be observed for the growth curves of the epidemic in the countries of Guinea, Sierra Leone and Liberia (Fig. 1). Most immediately, the epidemic in Liberia is growing at a much faster rate in Liberia than in Guinea. Although the epidemic likely began much earlier in Guinea ( [4]), Liberia had approximately the same number of cases in early August, twice as many cases by the end of August and nearly three times as many cases by mid-September. Even more striking, the number of cases in Guinea appears to have been growing sub-exponentially until late-August (approximately linearly with a slope of about 3 cases per day) while the number of cases in Liberia has been growing exponentially (approximately 10 cases per day averaged for July, 40 cases per day averaged for August and 70 cases per day in September). The growth dynamics of the epidemic in Sierra Leone appears to be intermediate between these two. It would be helpful to understand these different growth patterns within the context of a single epidemic, since a better understanding of the source of these different patterns may yield productive ideas for curbing the exponential growth of the epidemic in Liberia.

A variety of computational and statistical models have been used to help characterize and resolve the mechanisms underlying trends in the growth of this epidemic. The models of [5] and [6] include a parameter to estimate and predict the effect of control measures on the epidemic. SEIR models such as that of [5] and [7] are four-compartment models that resolve infectious dynamics between populations based on their susceptibility and infectiousness and account for the time scales of viral incubation and infectiousness. SEIR models with seven compartments ( [8, 9, 2]) further resolve the effects of varying degrees of transmission among, for example, community, hospital, and funeral populations.

These computational models focus on different aspects of the epidemic to explain or observe the marked differences of the growth curves for the epidemic in each country. The models [5] and [6], accounting for the effect of control measures, find that their models identify slowing of the growth of the epidemic compared to a free exponential only for Guinea and Sierra Leone. The model of [8] accounting for different community, hospital and funereal transmission rates predicts that a higher number of transmissions from funerals in Liberia could account for the faster rate of growth of the epidemic in Liberia compared to Sierra Leone. Likewise, the model of [2] predicted a higher fraction of patients in Liberia with no effective measures to limit transmission. Especially interesting differences among the three countries were described by [7] in their methodology to observe changes in the effective reproductive number over time. In particular they found that the effective reproductive number rose for Liberia and Guinea. The authors observe that this increase occurred somewhat early on during the Liberia outbreak in mid-July, when the outbreak spread to densely populated regions in Monrovia, and during the Guinean outbreak in mid-August, around the time the outbreak spread to densely populated regions in Conakry.

We would like to provide a proof of principle explanation for the differences in the dynamics of the 2014 Ebola epidemic based on differences in the community network properties of the affected regions, even as the number of daily interactions, transmission rates and in particular the average number of people infected by each person $R_0$ within a naïve population is the same for all three countries. A prediction of our model is that the effective reproductive number of the epidemic $R_e$ decreases quickly to values close to 1, indicating significant saturation effects for all three countries. Although we use a very simple homogeneous network model and our four model parameters are under-constrained, a particular choice of reasonable model parameters captures relevant average behaviors of the epidemic (for example, by fitting the growth curves of the epidemic) and we use these to interpret trends in the epidemic growth over time and between countries. A description of the roles of model parameters can be used to make predictions about the future growth of the epidemic in the context of potential epidemic controls that modify these parameters.

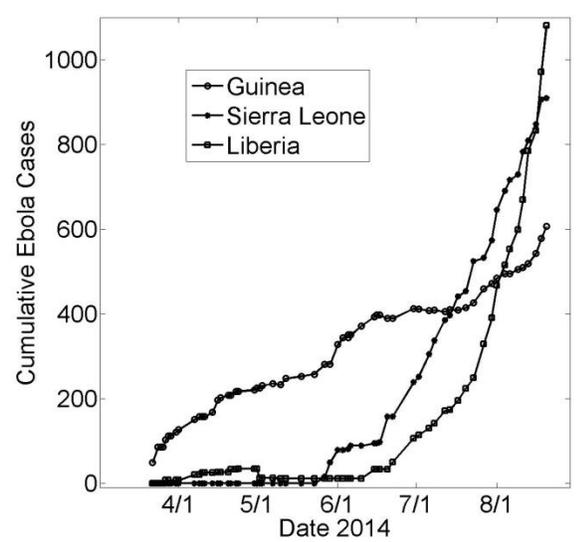

**Fig. 1: Cumulative Ebola Cases for Guinea, Sierra Leone and Liberia over Early 5 Months of the Epidemic (March 22$^{nd}$ – August 22$^{nd}$)**

Cumulative number of suspected or confirmed reported by the World Health Organization (WHO) March 22$^{nd}$ – August 22$^{nd}$. Five months of case data reports are measured since the initial outbreak report by the WHO on March 25$^{th}$ [10]. Case data was found on Wikipedia [11] compiled from WHO case reports [12] and was retrieved on October 15$^{th}$.

**The Importance of Network Interactions**
The roles of incomplete mixing within communities, heterogeneity in contact transmission, local saturation and the co-incidence of multiple transmission chains can have significant effects on epidemic dynamics ( [13, 14, 15]). The importance of network considerations, in particular the importance of reducing the contacts between exposed and unexposed groups, for controlling this epidemic has been described ( [16]). One of the most important features to capture about the Ebola virus (in particular for a model resolved at an individual, mechanistic level) is that there is a high probability of transmission between close contacts, but a lower probability among casual contacts. For example, in a formal study of transmission chains in the 1979 Ebola virus outbreak in Sudan, it was found that care-takers of the sick had a 5.1 higher rate of transmission than other family members with more casual contact ( [17]). Likewise, in the 1976 Ebola outbreak in Zaire, the probability of transmission was 27.3% among very close contacts (spouses, parents and children) but only 8% for other relatives ( [18]). In our model we organize individuals within households (a broadly defined term meant to represent the set of potential "close" contacts) and households are organized within local communities (a larger, modularly structured network of the population of less likely and less infectious interactions). The term "household" has been used in the literature for networks of close interactions, so we use this term here for convenience, but in the case of Ebola virus transmission, a network of close contacts would include overlapping household, hospital and funereal networks, and the concept of 'household' should be expanded to include these.

Two-scale community models with different transmission rates for close (or local) contacts and casual (or global) have been studied previously ( [19, 20, 21] and references therein) where the smaller scale compartment may be called households, clusters or sub-graphs. With this paradigm, there are two transmission rates, and thus two scale-dependent reproductive numbers that would sum to the global reproductive number $R_0$ of the epidemic. We define $R_{0H}$ as the average number of infections that occur within the household from an infected individual, and $R_{0C}$ as the average number of infections that occur within the community from a single infected individual. The reproductive number $R_0$ is the average number of infections expected in a completely naïve population (i.e., exactly one infected individual and no immunized individuals), so $R_0 \approx R_H + R_C$. However, in the context of local saturation effects, $R_0 = 1$ is no longer a relevant parameter for defining a threshold for epidemic growth ( [22, 23]).

We implement a discrete, probabilistic SEIR model on a social connectivity network that combines elements of a two-scale network model by [20] and the modular network of [24] to create what is technically a three-scale community network (individuals are organized within families, and families within modular local communities that are subsets of the entire population). This connectivity structure is simple by construction (more regular than small-world networks and lacking long-range connections of different lengths) yet nevertheless enables a parametrization of the level of mixing that exists between communities. Specifically, a population that is "well-mixed" has a larger number of families interacting within each local community. Details of the model are described in the Methods section. The effects of incomplete community mixing are significant even for moderate population sizes ( [19]) and cannot be reproduced by unstructured mean-field models that assume complete mixing (e.g., see [25]). This emphasizes the role of individual and network-based models to resolve these effects.

# Methods

A stochastic, individual-based SEIR model is implemented for a population with a network structure of two edge types: close contacts among members of a household and casual contacts among members of a local community. This component of the model is comparable to the two-scale SIR community model described by [20] (we use a discrete lattice-based simulation approach instead of a Markov approach and we add an exposed period for which an individual is exposed but not infectious). Local communities of each household are modeled as the set of $r$ nearest households, for this we refer to the modular lattice approach of [24]. See Fig. 2 for a schematic of our three-scale network. This last component enables us to systematically vary the extent of community mixing.

## Households of size *H*: Modeled on an $L \times H$ lattice

A population of size $P = L \cdot H$ is modeled as an $L \times H$ lattice where $H=|h_i|$ corresponds to a fixed number of individuals in a household $h_i (i = 1, \ldots, L)$. Each $i_{th}$ row of the lattice represents a distinct household $h_i$. Here the parameter $L$ is arbitrarily large so that $P$ represents an effectively infinite population (in simulations this only requires that $L \gg$ the number of households participating in the epidemic). and has a state $S \in \{S, E, I, R\}$ corresponding to whether the individual is susceptible, exposed, infected or refractory.

## Communities of size *C=2R+1* households: Modeled on an $L \times H$ lattice with modular structure

The population of households is further organized within communities $c_i$ $(i = 1, \ldots, L)$. The $i_{th}$ community $c_i$ is the set of individuals within all households within distance $r$ of the $i_{th}$ household (i.e., $c_i$ is the set of individuals within households $\{h_{i-r}, \ldots, h_{i-1}, h_i, h_{i+1}, \ldots, h_{i+r}\}$). Communities are comprised of a fixed number of households $(2r + 1)$ and thus each community is a set of $C \cdot H$ individuals. With periodic boundary conditions, a population with $L$ households would have $L$ overlapping communities. We choose a number of communities much larger than the number participating in the epidemic, so that our results are not affected by the choice of boundary conditions.

## Network Structure of Close and Casual Contacts on the Lattice

A network of two edge types representing close versus casual contacts are defined for the population (see Fig. 2). Individuals within a household are connected by edges that represent potential close contacts, and individuals within a community are connected by edges that represent potential casual contacts. Thusly, each household may be thought of as a well-mixed (completely connected) graph of *H* vertices and each community may be thought of as a well-mixed (completely connected) graph of *2r+1* vertices. From an individual's perspective, an individual located at the $(i, j)_{th}$ node is connected to each member of the $i_{th}$ household with potential close contacts and to each member of the $j_{th}$ community with potential casual contacts. The described network structure is completely homogeneous: every individual is centered within a network that is identical to that of every other individual.

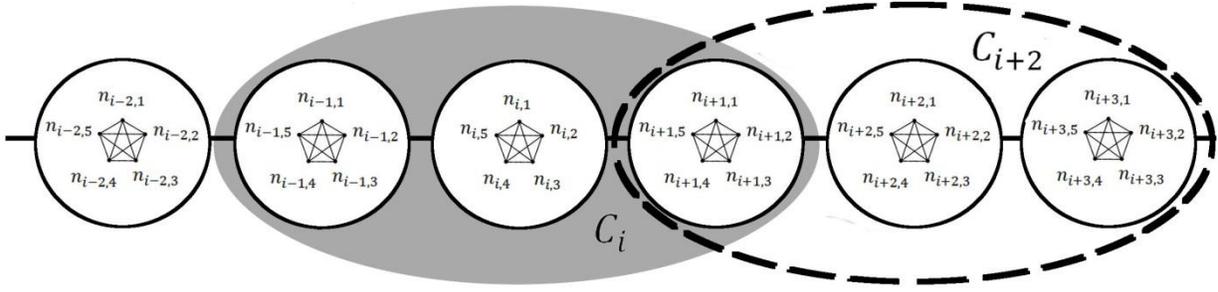

**Fig. 2: Network Structure of Individuals, Households and Two Communities $c_i$ and $c_{i+2}$**
The three scale community model with 5 individuals per household ($h=5$), and a community radius of $r = 1$ ($C=3$). Individuals $n_{i,j}$ within each household $h_i$ are completely connected by household-level contacts. Only two communities $c_i$ and $c_{i+2}$ are shown. All individuals in each community are completely connected by community-level contacts. Within the shaded region, the $C_{ith}$ community includes the three households $\{h_{i-1}, h_i, h_{i+1}\}$ while within the dashed region, the $C_{(i+2)th}$ community includes the three households $\{h_{i+1}, h_{i+2}, h_{i+3}\}$. Notice that there is some overlap between these two communities, as the two communities both contain household $h_{i+1}$. An infection in the $h_{ith}$ household would need to spread through the $C_{ith}$ community before the infections could spread to the $C_{(i+2)th}$ community. This figure was modified from Fig.1 of [20].

### SEIR Dynamics
Initially, all individuals (lattice nodes/network vertices) are susceptible (state **S**) except one individual that is exposed (state **E**) representing "patient 0". In simulations, time steps are discrete and correspond to exactly one day. States are updated at each time step with the following transition probabilities:

$p(S \rightarrow E) = probability\ that\ a\ susceptible\ will\ become\ exposed$
$\qquad = (1 - probability\ of\ no\ exposures\ from\ any\ infected\ contacts)$
$\qquad = (1 - (1-t_H)^{i_H} \cdot (1-t_C)^{i_C})\,,$

where $t_H$ is the transmission probability within a household (probability of exposure per day per infectious household contact), $i_H$ is the number of infectious household contacts, $t_C$ is the transmission probability within a community (probability of exposure per day per infectious community contact), and $i_C$ is the number of infectious community contacts. Transmission probabilities are the probability of transmission per day per infectious individual along a link between an infectious and susceptible individual, and are the products of (1) the probability that the two individuals interacting on a given day and (2) the probability that an interaction will result in exposure.

$p(E \rightarrow I) = probability\ that\ an\ exposed\ individual\ becomes\ infectious$
$\qquad = 1/\gamma,$
$\qquad$ where $\gamma$ is the average incubation period.

$p(I \rightarrow R) = probability\ that\ an\ infectious\ individual\ will\ become\ refractory$

$$= 1/\lambda,$$

where $\lambda$ is the average infectious period.

For the incubation and infectious periods, we follow recent modeling groups ( [5, 2] based on data in ( [26, 27, 28, 29]) using $\gamma = 5.3$ days and $\lambda = 5.61$ days. We observe that these periods may be longer ( [4, 29]), as used in ( [9, 6, 8]) or varied in ( [7]).

**Reparametrization of transmission rates in terms of $R_0$ and expected number of contacts η**
The extent of mixing increases with the size of the community $C$. A community of size $C$ means that each individual in that community has an equi-probable chance of interacting with each other member of the community. However, it becomes increasingly clear as the community size increases that members of the community will not interact with every member of the community every day. Rather, we assume that individuals interact with an average number of people η each day, where η is a fixed value independent of community size.

We define the reproductive numbers $R_{0H}$ and $R_{0C}$ as the average number of second infections resulting from a single infectious individual in a completely naïve (all susceptible) population due to household and community interactions, respectively. Since each network (household and community networks) is homogeneous, the values of $R_{0H}$ and $R_{0C}$ do not depend on the location of an infectious individual in the network and is proportional to the number of susceptibles in the network (equal to ($H$-1) or ($C$-1) in a naïve population) times the probability ($\lambda \cdot t$) that each susceptible individual will become exposed over the infectious period due to contact with the infectious individual:

$$R_{0H} \approx (H-1) \cdot \lambda \cdot t_H,$$
$$R_{0C} \approx (C-1) \cdot \lambda \cdot t_C, \text{[1]}$$

Since $R_0$ is a key epidemiological parameter, it is helpful to make this a control parameter of the model rather than the transmission rate. Thus, we solve for the transmission rates in terms of $R_{0H}$ and , or $R_{0C}$ and $C$: $t_C \approx \frac{R_{0C}}{(C-1)\cdot\lambda}$ $t_H \approx \frac{R_{0H}}{(H-1)\cdot\lambda}$, so that

$$p(S \to E) = \left(1 - \left(1 - \frac{R_{0H}}{(H-1)\cdot\lambda}\right)^{i_H} \cdot \left(1 - \frac{R_{0C}}{(C-1)\cdot\lambda}\right)^{i_C}\right).$$

**Four Free Model Parameters:** $R_{0H}, R_{0C}, H, C$.
Since our incubation and infection periods are fixed, our model is completely prescribed by four intuitive free parameters: the household reproductive number $R_{0H}$, the community reproductive number $R_{0C}$, the household size $H$ and the community size $C$. Note that since an infected individual becomes exposed from the community or from their household, we have $R_0 \approx R_{0H} + R_{0C}$, where $R_0$ is the usual reproductive number[2].

---

[1] Accounting for saturation effects that occur even over the infectious period of the first infectious individual in a naïve infectious populatiom, more accurate but less penetrable descriptions for the community network, for example, are given by

$R_{0C} = (C-1) \cdot \left(1 - (1-t_c)^\lambda\right)$ and $t_C = 1 - \left(1 - \frac{R_{0C}}{C}\right)^{\frac{1}{\lambda}}$.

[2] A better estimate would subtract the probability of a household contact exposed by both types of transmission: $R_0 = R_{0H} + R_{0C} - \frac{R_{0H} \cdot R_{0C}}{(C-1)}$.

**Model Response Parameters**

Our model tracks the states of individuals over time. In simulations, an individual is defined as an Ebola case when they become infectious. This assumes that an individual is not recognized as a case until they are infectious and that there is no delay in identifying infectious individuals.

Many response variables can be calculated, included the fraction of infections not occurring due to a contact already being infected (saturation), the rate of spread of the infection through the population (as, for example, a function of the average distance from the initial infected individual), the structure of the chain of transmission from any single exposed individual, etc. In the results presented here, we focus on the number of cases per day. We also calculate the effective reproductive number $R_e$ which is the average number of infections resulting from each infectious individual. At the conclusion of a simulation, $R_e$ is calculated as the total number of infections caused by any refractory individual divided by the total number of refractory individuals (i.e., care is taken to not include the data of infectious individuals that have not yet become refractory.)

**Case Data and Matching Epidemiological Curves for Ebola Case Number versus Time**

We compare simulated Ebola cases per simulation days with Ebola cases per day for Guinea, Sierra Leone and Liberia. Case data was found on Wikipedia [11] compiled from WHO case reports [12] and was retrieved on October 15$^{th}$.

Since the date of the first case is not given, and especially since the number of days between the first case and the $n_{th}$ case can be highly variable, we synchronize the simulation day and the calendar date by using the first day that there are 48, 95 or 51 cases in Guinea, Sierra Leone and Liberia, corresponding with March 22$^{nd}$, June 15$^{th}$, and June 22$^{nd}$, respectively. In Figs. 3 and 6, simulation total case numbers are shown as the average results of $n=100$ simulations ± the standard error (the standard error is calculated as the standard deviation divided by $\sqrt{n}$). A simulation is only counted and included in the 100 replicates if the outbreak does not die out before reaching the observed number of cases.

To model the epidemics of Guinea and Liberia after August 22$^{nd}$, we initialize outbreaks in new populations and model the case counts over time with their sum. We use a systematic procedure to determine the time points at which to introduce a new outbreak. We introduce an $n_{th}$ population with 100 cases on the calendar day the total cases predicted by the first (n-1)$_{th}$ outbreak populations falls short of the actual data by 100 cases.

Simulations were completed in the software package Matlab, and Matlab scripts used to generate all figures can be found at:

http://www.southalabama.edu/mathstat/personal_pages/byrne/PLoS_MATLABscripts.htm

# Results

**Fitting Country Case Data for the First Five Months of the Epidemic**
The number of cases over time for Guinea, Sierra Leona and Liberia over the first five months of the epidemic (March 22$^{nd}$ – August 22$^{nd}$) is shown in Fig. 3. We sought to describe these dynamics in the context of the spread of a single epidemic within a contiguous region, so that key epidemiological parameters, in particular transmission rates and the basic reproductive number, would be the same in each country. We also assumed average household size would be the same for each country and required that the difference between countries be described solely by changing the community size, $C$, a measure of the size of the community within which infectious and susceptible individuals interact by casual contacts. It is thus natural to refer to $C$ as the *community mixing size*.

For the fixed set of parameters $H=16$, $R_{0H}=1.8$ and $R_{0C}=0.55$, average simulation results yielded good fits[3] to the empirical increase in case number over time as the community mixing size $C$ was increased from $C=9$ to $C=25$ to $C=51$ for Guinea, Sierra Leone and Liberia (Fig 3). This result provides a *proof of principle* that the differences in the growth dynamics of the epidemic can be explained by different levels of community mixing.

The particular set of values in Fig. (3) were chosen to be within likely ranges of their empirical values (see Discussion). A local sensitivity analysis was done to establish that $R_{0C}$ and $C$ were locally optimized for the given choice of $H=16$, $R_{0H}=1.8$, but a systematic search of parameter space was not done to determine the shape of the set of all parameters, still within the likely range of their empirical values, that would provide equally good fits of the data. Since the epidemic growth rate generally increases with increases of any one parameter, different parameter values may yield similar results if one parameter is raised while decreasing another, though we note there are complex effects on the shape of the transient behavior in each case. In the Appendix, we include several examples of alternative fits to the data in which, for example, the household size $H$ is decreased or increased, or the household reproductive number $R_{0H}$ is increased or decreased. These changes to $H$ and $R_{0H}$ require commensurate changes in $C$ and $R_{0C}$ in order to achieve good fitting of the data. In these results, we focus on the parameter set $\{H, R_{0H}, R_{0C}\}=\{16,1.8,0.55\}$ since these are consistent with previously described epidemiological data for Ebola virus (see discussion for further details).

The effective epidemic reproduction number $R_e$ is calculated from the model simulations. The value of Re is time-dependent (plots are provided in the Appendix) but decrease to values close to one for each country ($R_e$=1.03±0.01, 1.10±0.02 and 1.17±0.02 for Guinea, Sierra Leone and Liberia, respectively). These values indicate very strong saturation effects, as many members of a household do not infect anyone after all members of their household have been exposed.

---

[3] *R*-squared values will be reported in the revision.

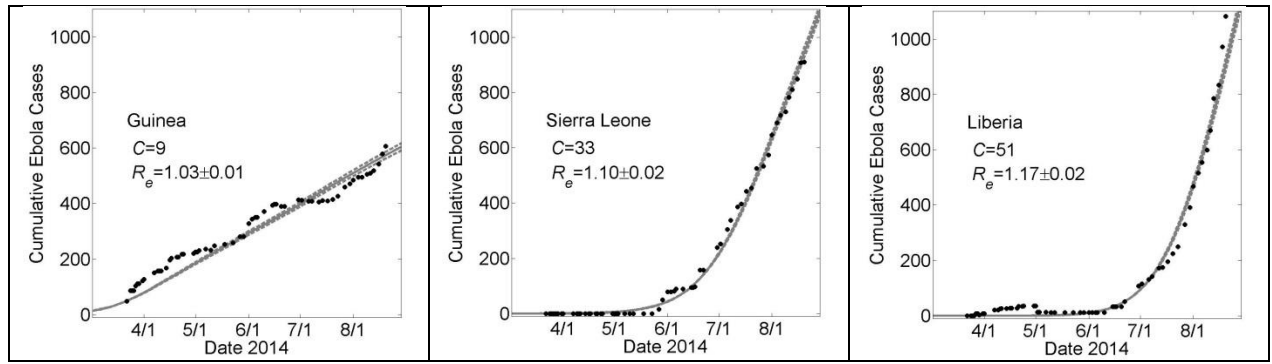

**Fig. 3: Country Specific Model Fits During 5 Months 3/22 – 8/22 of the 2014 Epidemic**
These graphs show average simulated cumulative number of cases (gray solid curve) for the community size *C* that best fits the reported WHO data (filled black circles) from March 22, 2014 to August 22, 2014 while all other parameters were held constant. For all simulations, the household size $H$=16, $R_{0H}$=1.8 and $R_{0C}$=0.55. For Guinea, Sierra Leone and Liberia, best fits were found for community sizes $C$={9,33,51}, respectively. Simulation results of cumulative Ebola cases results are shown as mean ± standard error. The effective reproductive number $R_e$ on the simulation day corresponding to August 22$^{nd}$ is provided as mean ± standard deviation. In Fig. 6, the parameters generating these fits will be referred to as "Guinean-type outbreak parameters", "Sierra-Leonean-type outbreak parameters", and "Liberian-type outbreak parameters", respectively.

**Stochastic Variability of Individual Simulations and Stochastic Spread of the Epidemic**
The simulation curves in Figs. 3 and Fig. 7 show the average number of Ebola virus cases versus time for simulations that did not die out before reaching a threshold number of cases. Since our model is probabilistic, individual simulation curves are variable (See Fig. 4a) and the epidemic fails to reach the threshold number of cases in at least 45% of the simulations due to the infection failing to leave the household (0.45=1-$R_{0C}$). Since the basic reproductive number is not predictive of the spread of an epidemic in clustered networks ( [23, 30]), and since stochastic effects are significant (Fig. 4), we describe the likely spread of an epidemic with a histogram of the distribution of outbreak sizes in Fig. 4*bcd*. The distribution of outbreak sizes for our model parameters describing the growth dynamics of Guinea and Liberia (Fig. 4*cd*) are biphasic: simulations frequently result in no spread (the epidemic fails to spread in the population if it fails to spread in the household ( [31])) or is likely to becomes quite large (>900 infections) if it does spread. For Guinea, there is some non-negligible probability of small outbreaks that do not become epidemics (Fig. 4*c*). For the parameters that describe Liberian growth dynamics (Fig. 4*d*), the probability of intermediate sized outbreaks is negligible (<1/900). To underscore the possibility that Guinea is a population that is in transition from one that would likely have only small outbreaks to one that would always have large epidemics like Liberia, we provide a histogram of the distribution of outbreak sizes for a population with even smaller community mixing size (*C*=5) (Fig. 4*b*). For this population, there is a sizable probability of small outbreaks, with a negligible probability of a large one.

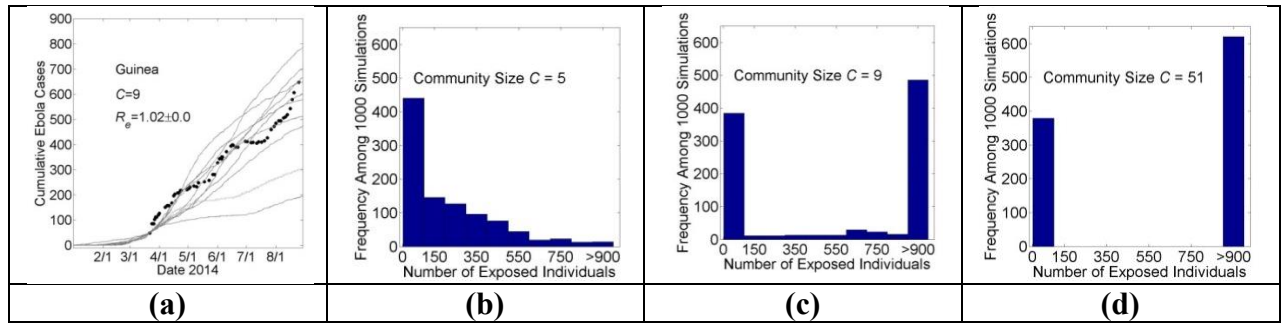

**Fig. 4: Probabilistic results of the model.**
In panel (a), the results of are shown for 10 simulations of Guinean-type outbreaks ($H$=16, $R_{0H}$=1.8, $R_{0C}$=0.55, $C$=9, see Fig. 3) that persisted for at least 1000 exposed individuals. In the remaining three panels, histograms are shown of the frequency of outbreaks of size $n$ (where the outbreak size is defined as the total number of individuals exposed during the outbreak) for 1000 simulations as the community mixing size was varied from $C$=5 (putative historical-type outbreak), $C$=9 (Guinea-type outbreak), and $C$=51 (Liberian-type outbreak).

## Location of Country Parameters in Phase Space and Predictions for Curbing Epidemic Growth

The locations of parameter values in phase space that fit the Guinea, Sierra Leone and Liberia data are shown in Fig 5. in Fig. 5a, the community reproductive number $R_{0C}$ is varied along the *x*-axis from 0.05 to 0.85, and the community mixing size $C$ is varied along the *y*-axis from 1 to 65. The three countries are located along the *x*-axis at $R_{0C}$=0.55, and along the *y*-axis at $C$=9 (Guinea), $C$=25 (Sierra Leone) or $C$=51 (Liberia). The shaded regions indicate increasing numbers of cases in the first 100 days after the outbreak has been established (defined by reaching 50 cases). This diagram shows the effect of decreases in $C$ and $R_{0C}$ on the growth of the epidemic during an early time period. For example, the phase diagram of Fig. 5a predicts that Liberia could have the slower Guinea-type growth dynamic if the reproductive number within communities was decreased from $R_{0C}$=0.55 to $R_{0C}$=0.25 (approximately a 50% reduction in community transmission).

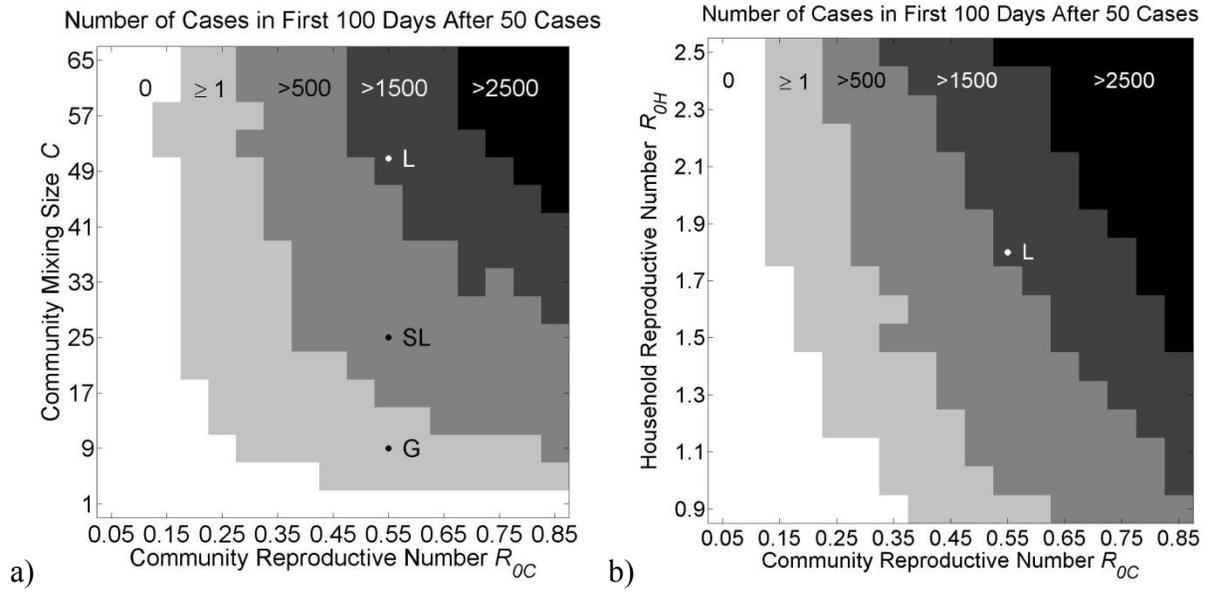

**Fig. 5: Phase Diagrams of Parameter Space**
Each phase diagram shows the growth of cases in the first 100 days after the outbreak is established (when the total number of cases exceeds 50 cases). Panel (a): In the first diagram, the community mixing size $C$ and the community reproductive number $R_{0C}$ are varied as all other model parameters are held fixed ($H=16$, $R_{0H}=1.8$). The locations of parameters used to fit the growth curves for Liberia (L), Sierra Leone (SL) and Guinea (G) in Fig. 3 are indicated along the vertical line $R_{0C}=0.55$. ). Panel (b): In the second diagram, the household reproductive number $R_{0H}$ and the community reproductive number $R_{0C}$ are varied as all other model parameters are held fixed ($H=16$, $C=51$). The locations of the parameters fitting the growth curve for Liberia is indicated at the cross-section of the vertical line $R_{0C}=0.55$ and the horizontal line $R_{0H}=1.8$. In both panels, the unshaded white areas are the regions of phase space where fewer than 1% of simulations resulted in 50 cases.

**Extrapolating from the Five Months March 22$^{nd}$ – August 22$^{nd}$ to Interpret Epidemic Dynamics through October 15$^{th}$**
During the writing of this manuscript, the growth dynamic of the epidemic in Guinea changed abruptly by an increase in the average number of Ebola cases per day. Our model parameters that provided a good fit to the Guinea data March 22$^{nd}$ – August 22$^{nd}$ (Fig 3a) predicted that the number of cases in Guinea would continue with its previous trend of approximately 3 new cases per day (Fig 6a, the solid black line). However, the abrupt change in slope suggested a new source of cases. We calculated the difference between the actual number of Ebola cases and the predicted number of Ebola cases over time to find the number of 'new cases' that would be supplied by a putative second outbreak. In particular, there were 823 cases reported for Guinea on September 3$^{rd}$ whereas our model predicted only 623. We thus used our model to fit a second Guinean outbreak with 200 cases on September 23$^{rd}$ (dotted line). The sum of the two outbreaks fit the data for Guinea very well when the community mixing parameter was $C=51$ for the second Guinean outbreak. Our model predicted that this second outbreak began with its first case in early July.

Likewise, our model parameters that provided a reasonably good fit for the growth of the epidemic in Liberia over the earlier time period made a poor prediction for the growth over the next six weeks (Fig. 6b). Our model generally predicts a growth dynamic that is linear after a transient exponential period. We repeated the method described for Guinea for fitting the Ebola cases that were in excess of our modeling predictions and these results are summarized in Figs. 6b and 6c.

In Fig 6b, a guiding exponential is drawn (dashed line) to show that the cumulative case data for Liberia is fit well by an exponential. Since early case data is quite noisy, we used the exponential fit to predict that the first 100 cases would have occurred on July $10^{th}$. The predictions of simulations, in which the day of the $100^{th}$ case was defined to be July $10^{th}$, are shown in Fig. 6b (solid gray line). Simulation predictions are indistinguishable from the exponential through mid-August, confirming that simulations predict a growth dynamic that is transiently exponential. After this transient exponential period, the predicted growth dynamics was linear (with slope of approximately 30 cases per day).

We repeated the procedure described for fitting the Guinean case data, in which a new outbreak with 100 cases is simulated each time the number of reported cases exceeds the number of predicted cases by 100 cases. This method predicted three independent outbreaks, all with community mixing size C=51, beginning in late June, mid-July and late July. While the superposition of these simulation outbreaks resulted in a very good fit of the data, especially the transient growth dynamics, we note that equally good fits could be certainly achieved by other linear combinations of increasingly smaller outbreaks.

Generally, our model predicts a growth dynamic that is linear after a transient exponential growth period. In the context of our modeling framework, without the simulation of secondary outbreaks, the linear growth can be understood as the depletion of susceptibles within participating communities, so that new infections occur only with the exposure of new communities at the boundary of already exposed communities. The slope of this increase will depend in a complex way on the propagation of saturation effects from infected sources and on the connectivity of individuals between adjacent communities.

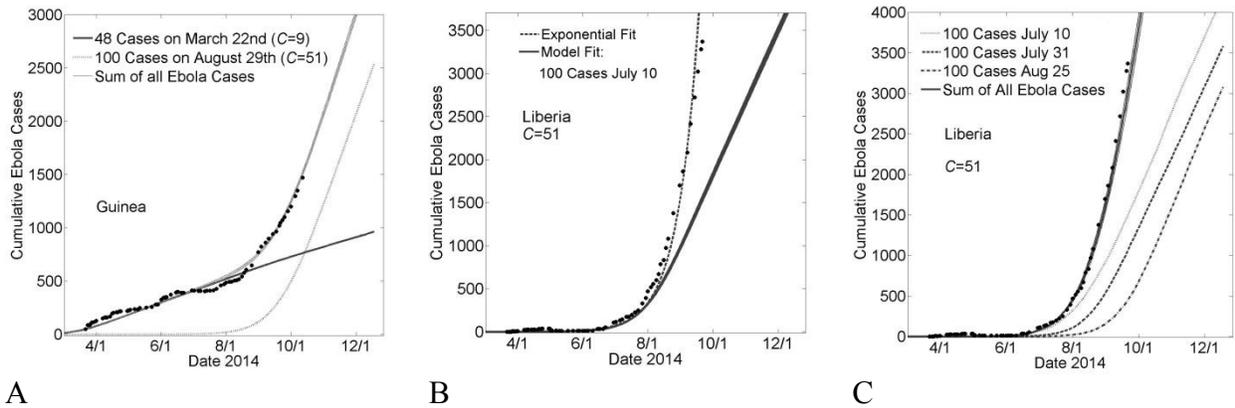

A    B    C

**Fig. 6: Country Specific Model Fits for the Sixth Month of the Epidemic (Guinea and Liberia)**

**Panel A: Modeling the increase in slope of Guinean Ebola cases with the seeding of a second outbreak** Model parameters for a Guinean-type outbreak (H=16, R0H=1.8, R0C=0.55, C=9, see Fig. 3) predict a continued linear increase in cases (solid line '48 Cases on March $22^{nd}$'). A second outbreak is modeled with Liberian-type outbreak parameters (H=16, R0H=1.8, R0C=0.55, C=51, see Fig. 3) with 100 cases on August $29^{th}$ (dotted line). Their sum is provided as a model fit of the Guinean Ebola case data for the sixth month of the epidemic ('Sum of all Ebola Cases', shown as mean ± standard error).

**Panel B: Model prediction for Liberia is transiently exponential and then linear increase in Ebola Cases** Early data for Liberia is particular noisy, but a guiding exponential shows that the growth of cases in Liberia is approximately exponential (dashed line). The guiding exponential is used to estimate that the first 100 cases occurred on July $10^{th}$. Fitting this constraint, Model parameters for a Liberian-type outbreak (see Fig. 3) predict a linear increase in cases after a transient period of exponential growth (thick gray line, 'Model Fit').

**Panel C: Modeling the continued exponential growth for Liberian Ebola cases with three independent seeding events** Three outbreaks are modeled with Liberian-type parameters (H=16, R0H=1.8, R0C=0.55, C=51, see Fig. 3) with 100 cases on July $10^{th}$, July $31^{st}$ and August $25^{th}$ respectively. Their sum is provided as a model fit of the Guinean Ebola case data for the sixth month of the epidemic ('Sum of all Ebola Cases', shown as mean ± standard error).

## Discussion

Fitting the growth dynamic of the epidemic for each country over the five months March 22$^{nd}$ – August 22$^{nd}$ of the epidemic while varying only the community mixing size *C* provides a *proof of principle* that accounting for local effects such as saturation and heterogeneous transmission among contacts can account for the differences in the rate of growth of the epidemics in different countries, and that they enable us to view qualitatively different growth trends as different facets of a single epidemic.

### Confidence in and Interpretation of Model Parameters

Our model had three parameters ($R_{0H}, R_{0C}, H$) which needed to be estimated to constrain the model and a fourth parameter (the community mixing size, C) which was varied to fit the different trends in case data over time for Guinea, Sierra Leone and Liberia. The reproduction numbers within households $R_{0H} = 1.8$ and within communities $R_{0C} = 0.55$ were constrained by the observation that $R_0 \approx R_{0H} + R_{0C}$, consistent estimates by several different groups that $R_0$ is between 1 and 3 ( [5, 4, 6, 9]), and a study of transmission chains in the 1979 Ebola virus outbreak in Sudan that found that care-takers of the sick had a 5.1 higher rate of transmission than other family members with more casual contact ( [17]). A household size of *H*=16 individuals is much larger than the actual average West African household size of 5 to 6 (e.g. [32]). However, the household grouping in our model is intended to describe the set of individuals that might potentially share close contacts, and is thus conceptually extended to include many different kinds of groupings within which close contacts with an infected individual might be expected, such as extended family networks (multiple family units are frequently found in West Africa ( [33])), overlapping with networks funereal preparations and rituals, and health care worker networks (e.g., nurses sharing shifts with a sick patient at a hospital ( [34]), etc . Considering these extended groupings, an average household size of 16 individuals was chosen. Also, a local sensitivity analysis found that this parameter value was optimal given the specified household and community reproductive numbers $R_{0H} = 1.8$ and $R_{0C} = 0.55$.

For the parameter values ($R_{0H} = 1.8, R_{0C} = 0.55, H = 16$), the model predicted community mixing sizes of *C* = 9, 25 and 51 fitting the case number over time for Guinea, Sierra Leone and Liberia for the five months March 22$^{nd}$ – August 22$^{nd}$ of the epidemic (Fig 3). These parameter values show that the difference in Liberia and Guinea can be explained by a 5-fold difference in their community mixing sizes. One description of this result is that if their social network structures are quite similar, so that all populations are composed of many communities of the smaller mixing size *C*=9, the large community mixing size *C*=51 can be interpreted as a threshold number of interactions between communities that effectively blur, on average, the boundary between communities on a scale of about five communities. In other words, this difference can be quite subtle. Community mixing sizes are relatively small ($H \times C =$ 144 to 816 individuals) compared to the size of the local population in towns or cities, so saturation effects within these communities are significant well before the susceptible population in the local community is depleted.

Our model provides a proof of principle that the differences in growth dynamics in different regions of West Africa can be explained by differences in community mixing sizes. This should not be interpreted as a prediction that there are not significantly different levels of epidemic control in the different regions. On the contrary, effective and ineffective epidemic controls can have significant effects, positively or negatively, respectively, on the community mixing size. In particular, similar to households [31], hospitals can have an amplifying effect by encouraging transmission among patients and through health care workers ( [18]), or can limit community mixing by isolating patients. Indeed, while isolation may be modeled as decreasing the transmission rate between infected and susceptible individuals, it is arguably a more accurate description that the transmission rate per contact is fixed, but the population of potential contacts is greatly reduced. In contrast, measures such as frequent hand washing and chlorine stations reduce the transmission rate without affecting community mixing size. Since our model assumes constant transmission rates throughout the region, the effects of such measures would be (erroneously) averaged within apparent differences in the community mixing size.

We found that to model increases in the epidemic growth rate in the sixth month of the epidemic, we needed to model the spread of the epidemic into new communities. While this might initially be interpreted as a failure of our model to predict the epidemic past the first five months of the epidemic to which we fit our parameters, on the contrary relatively constant trends over the first five months enabled us to determine reasonable parameters while the subsequent spread of the epidemic into new communities eventually, as in the sixth month, is not only intuitive but eventually inevitable in an as of yet uncontrolled epidemic. Previous epidemiologists ( [35] and [23]) have reasoned that an epidemic over a large geographical region is best considered as the superposition of many smaller outbreaks, with transmission and saturation effects occurring at this local level, with occasional long-range interactions driving the spread across the region.

**The Importance of Community Mixing Size and Saturation Effects**
Our individual-based model accounting for saturation effects was able to fit early outbreak data for Guinea, Sierra Leone and Liberia by changing one community parameter, the community mixing size $C$, while holding other important epidemiological parameters constant. Potential saturation effects in the empirical data may be observed by the difference in the case data curves from exponential; Guinea had a linear growth trend in cases over the first five months of the epidemic, while Liberia was best fit by a simple exponential [6]. Heuristically, saturation can be understood by considering the extent to which limited community mixing results in overlapping (and redundant) chains of transmission as two infected families have a higher probability of being located within the same mixing community, and infecting families within the same community, than two randomly chosen families ( [15] )

Saturation effects are reflected in the model by differences in the effective reproductive number $R_e$ and the basic reproductive number $R_0$, and especially by decreases in $R_e$ over time. Without saturation effects, $R_e$ and $R_0$ will be equal; however as saturation effects increase $R_e$ approaches 1. Our model predictions for $R_e$ decreased to values close to 1, indicating significant saturation effects. Especially, saturation effects within households will reduce $R_e$ towards 1 since the household reproduction number $R_{0H}=2$ is the main contribution to the basic reproduction number $R_0$ =2.35 and most infected individuals within households will find that some members of their households are already exposed. Note that saturation effects cannot drive $R_e$ below 1. (When saturation effects are maximized, everyone is infected and $R_e$ =1.)

## Basic Reproductive Number $R_0$

The moment that an individual becomes infectious in an entirely naïve community, they will on average infect a number of people $R_{0H}$ in their household and a number of people $R_{0C}$ in their community, so that $R_0 \approx R_{0H} + R_{0C}$. By fitting the epidemic dynamics for $R_H = 1.8$ and $R_{0C} = 0.55$, our model demonstrates that an $R_0$ of 2.55 is consistent with this epidemic. For the commonly understood definition of $R_0$ that we use here, in which the community is entirely naïve, it is important to emphasize that the infection is the result of a one-off interaction of an individual with a distant community, or a first zoonotic event, since the expected prior distribution of infections around an infectious individual, even at very early times within an epidemic, will change within the course of a single infection. The basic reproductive number $R_0=1$ in this context does not provide a threshold for the growth of the epidemic. In well-mixed population models, $R_0=1$ is the threshold for an epidemic. It is clear however, that the basic reproductive number loses this prediction capacity in the context of network models, since even if the spread of infection is very high in households ($R_{0H} \gg 1$), the infection will not spread if $R_{0C}$ is so low that the infection cannot leave a household. Further, a relatively low community transmission rate, can cloak a relatively high household reproductive number (see Appendix 2 for such an example). Clustering of individuals within households can act as an amplification center and stabilize spread of an epidemic, or contribute to saturation effects that inhibit spread ( [31]). The limitations of this reproductive number in predicting the course of an epidemic can be addressed to some extent by descriptions of the household-to-household reproductive number (the average number of households infected by an infected household) (e.g. [21]).

## Model Predictions for Limiting Exponential Growth of the Epidemic in Liberia

Our model provides a proof-of-principle that sub-exponential growth is possible, and indeed explains the linear growth of the epidemic in Guinea over the first five months of the epidemic, if community mixing is limited. Epidemic control measures can have a variable effect on community mixing sizes. Our model results indicate that epidemic control measures can be further optimized by considering implementations that decrease community mixing. Optimal ethical and pragmatic epidemic control measures are beyond the scope of this manuscript. Instead, we generalize that considered epidemic control measures will work by decreasing transmission by decreasing transmission rates and/or decreasing community mixing rates (e.g., chlorine protocols, quarantines, isolation, etc.) The phase diagrams in Fig. 5 shows the location of Liberia, Sierra Leone and Guinea in $C$-$R_{0C}$ and $R_{0H}$-$R_{0C}$ phase space (monotonically mapping to differences in community mixing size and community transmission rates) and movements within this space show the effect on the number of cases from 50 infectious individuals over a particular 100 day period.

## Differences in Ebola Virus Epidemic Growth Dynamics over Time and Geographic Region

Our model results show that the spread of an epidemic due to the introduction of an infected individual within the community is not inevitable since even for community with relatively large community mixing sizes, there is a significant probability that the infection fails to spread from one infected individual. However, our model also shows that as the community mixing size increases, once the epidemic begins spreading, it becomes increasingly inevitable that the outbreak will be very large. Similar dependence of the distribution of outbreak sizes were observed in small world networks ( [15]). Our model prediction that communities with small mixing sizes will have a distribution of smaller outbreaks is consistent with the historical observation of an outbreak in 1976 that the Ebola virus did not seem to spread well and no transmission chains longer than three were found in the outbreak ( [17]). A recent study estimated that the probability of a large outbreak of more than 1,000 cases in 1976 was 3% [36]. If community mixing sizes increase with historical time, which seems likely with increased population density and urbanization, then the probability of very large outbreaks increases with time and Guinea may be an example of a population that is in transition between a population where small outbreaks would occur versus large epidemics. As noted in ( [36]), these trends would be further exacerbated by concurrent increases in interactions between humans and animal carriers as humans spread into new animal habitats.

Focusing on the dynamics of the epidemic in the five months March $22^{nd}$ – August $22^{nd}$ might have suggested that Guinea and Liberia were quite different in some aspect, since the epidemic was not growing exponentially in Guinea but was growing exponentially in Liberia. However, an abrupt change in the growth dynamic of Guinea, and our model prediction that the Liberian growth may be best described as the superposition of several outbreaks staggered in time, suggest that the specific early dynamics of the outbreak may be probabilistic, depending on the characteristics of community where the outbreak begins. As the epidemic spreads to new communities, the growth dynamics represent the sum of these contributions and are dominated by the fastest growing outbreaks. Our model predicts that in a single community outbreak, there is a transient period of exponential growth followed by a linear increase in cases. This reflects the wave-like spread of the infection and has been previously observed previously in small-world networks ( [15]).  While our model predicts that each individual outbreak will saturate, the epidemic will remain exponential if the virus continues to seed in new communities. Likewise, it was observed that even a small number of long-range links in a small-world network results in a dramatic increase in the growth rate of the epidemic ( [15]).

**Model Scope, Limitations and Future Directions**
Our three-scale model is a model of intermediate complexity that accounts for heterogeneous transmission and, like more complex models such as small-world networks ( [15]), permits systematic variation of the extent of community mixing (modeled with one other network scale). The model results predict a household size 2-3 times larger than a typical household and thus this cluster likely represents a connected network of households and other groupings where close contacts would be expected. This indicates that a hierarchical network model might be more appropriate. Our network choice was preliminarily simplistic, and other types of networks, such as hierarchical and scale-free networks, are better descriptions of community connectivity in real-world populations. In particular, simulating the 'seeding' of the outbreak in new communities as a discrete event is only appropriate for a small number of events. Small-world networks with a heterogeneous distribution of edge lengths (many short-range interactions and a small number of long-range interactions) would be able to model seeding events probabilistically, without requiring explicit inputs for when these events occur. Modeling the Ebola virus epidemic in the context of different network formulations could verify that these described results are not dependent on specific choices of model implementation. Since our model parameters were not fully constrained, virtually any individual-based data, such as the average fraction of immediate family members that become ill, or information about the distribution of the number of infections resulting from infected individuals, can be used to constrain these parameters. Further, a network model with additional spatial information could be used to further explore the effectiveness of various epidemic control strategies, as the spatial movement of the epidemic through new communities, and resurgence through previously exposed communities, will have profound effect on the persistence of the epidemic ( [15, 37, 38]).

**Conclusions**
Our model demonstrates that sub-exponential growth is a possible long-term trend if community mixing is limited, and provides predictions on how changes in transmission rates will result in decreases in the growth of the epidemic. Our results suggest that limiting community mixing over this scale would be an important consideration while designing epidemic control strategies. Community mixing sizes consistent with model fits of the data are quite small, with 100 to 1000 individuals. It is especially important to prevent the seeding of the outbreak into new communities, as the diffusion of transmission chains into new communities is what maintains the exponential growth of the epidemic.

# Appendix 1: Effective Reproductive Number $R_e$

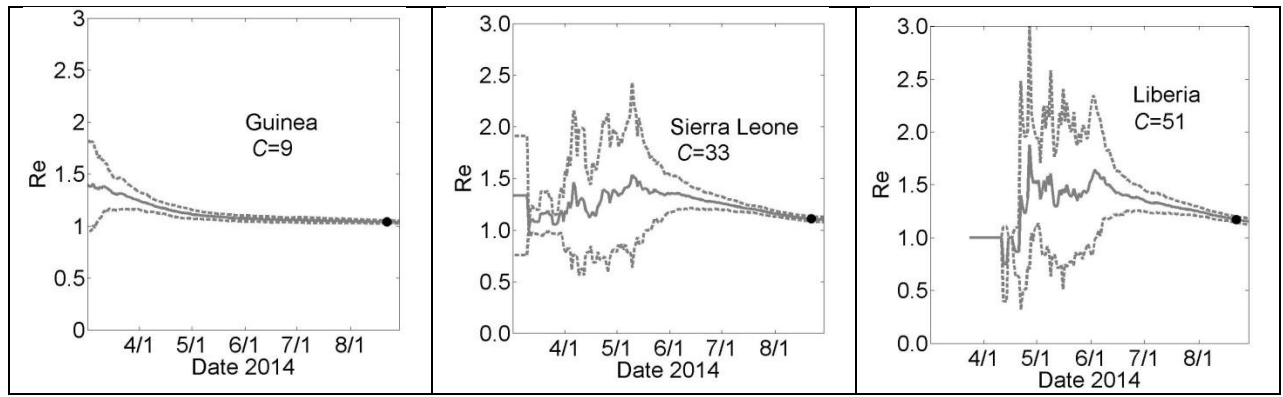

**Figure Appendix 1**: Country Specific Effective Reproductive Number $R_e$ as a Function of Time for the simulation results described in Fig. 3.

These graphs show the calculated effective reproductive number $R_e$ for Guinea, Sierra Leone and Liberia for the simulations described in Fig 3. Results are shown as mean (solid line) ± standard error (dashed lines).

# Appendix 2: Simulation Fits of the Data for Other Sets of Parameter Values

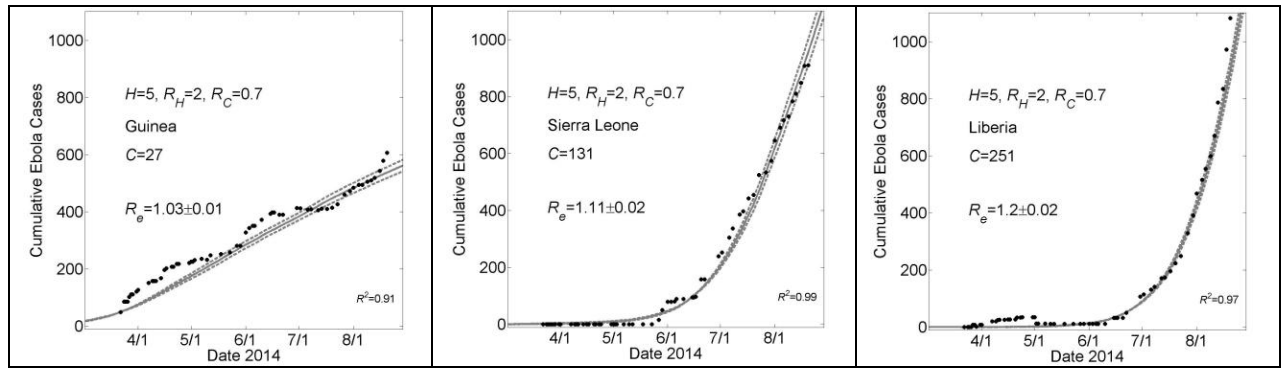

**Figure Appendix 2A: Country Specific Model Fits During 5 Months March 22$^{nd}$ – August 22$^{nd}$ of the 2014 Epidemic For a Lower Value of *H***
These graphs show simulated cumulative number of cases (gray curves) for the community size *C* that provides a fit of the reported WHO data (filled black circles) from March 22, 2014 to August 22, 2014 while all other parameters were held constant. For all simulations, the household size *H*=5 and the basic reproductive numbers within households and within communities were respectively $R_H$=2 and $R_{0C}$=0.7. For Guinea, Sierra Leone and Liberia, fits were found for community sizes *C*={27,131,251}, respectively. Simulation results of cumulative Ebola cases results are shown as mean ± standard error. The effective reproductive number $R_e$ on the simulation day corresponding to August 22$^{nd}$ is provided as mean ± standard deviation.

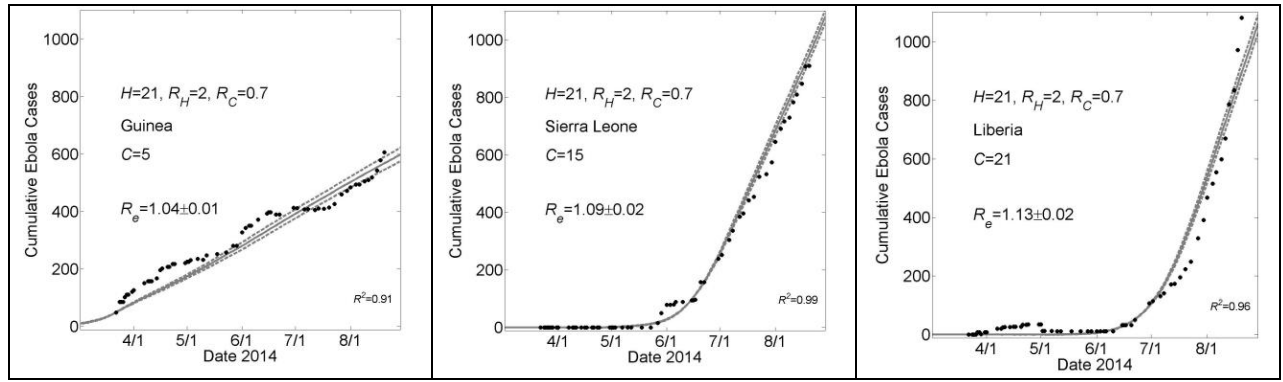

**Figure Appendix 2B: Country Specific Model Fits During 5 Months March 22$^{nd}$ – August 22$^{nd}$ of the 2014 Epidemic For a Higher Value of *H***
These graphs show simulated cumulative number of cases (gray curves) for the community size *C* that provides a fit of the reported WHO data (filled black circles) from March 22, 2014 to August 22, 2014 while all other parameters were held constant. For all simulations, the household size *H*=21 and the basic reproductive numbers within households and within communities were respectively $R_H$=2 and $R_{0C}$=0.7. For Guinea, Sierra Leone and Liberia, fits were found for community sizes *C*={5,15,21}, respectively. Simulation results of cumulative Ebola cases results are shown as mean ± standard error. The effective reproductive number $R_e$ on the simulation day corresponding to August 22$^{nd}$ is provided as mean ± standard deviation.

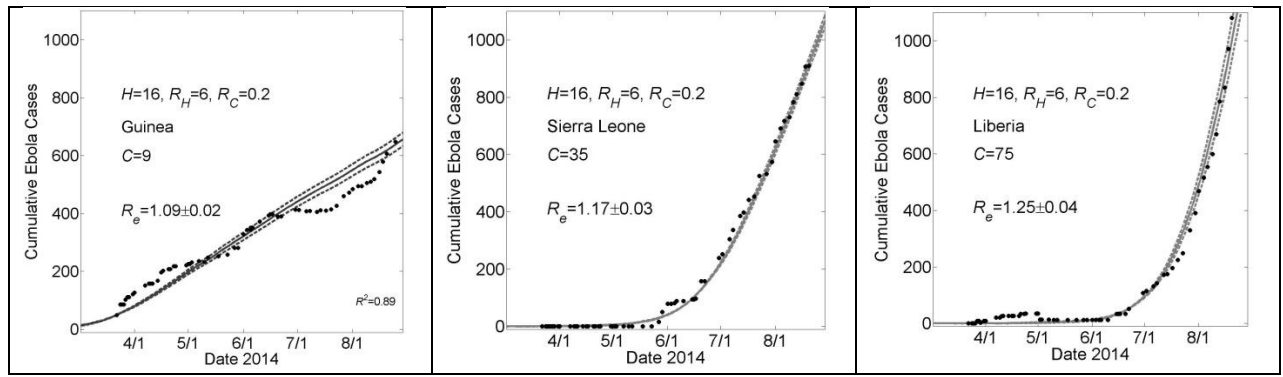

**Figure Appendix 2C: Country Specific Model Fits During 5 Months March 22$^{nd}$ – August 22$^{nd}$ of the 2014 Epidemic For a Higher Value of $R_H$**

These graphs show simulated cumulative number of cases (gray curves) for the community size $C$ that provides a fit of the reported WHO data (filled black circles) from March 22, 2014 to August 22, 2014 while all other parameters were held constant. For all simulations, the household size $H=16$ and the basic reproductive numbers within households and within communities were respectively $R_H=6$ and $R_C=0.2$. For Guinea, Sierra Leone and Liberia, fits were found for community sizes $C=\{9,35,75\}$, respectively. Simulation results of cumulative Ebola cases results are shown as mean ± standard error. The effective reproductive number $R_e$ on the simulation day corresponding to August 22$^{nd}$ is provided as mean ± standard deviation.